\PassOptionsToPackage{colorlinks=true, linkcolor=blue, citecolor=blue, urlcolor=blue, bookmarks=true, bookmarksnumbered=true, pdfstartview=FitH}{hyperref}
\documentclass[runningheads]{llncs}
\usepackage[T1]{fontenc}
\usepackage{graphicx}
\usepackage{enumitem}
\usepackage{marvosym}
\usepackage{dblfloatfix}
\usepackage{float}
\usepackage{afterpage}
\usepackage{svg}
\usepackage{transparent}
\usepackage{array}
\usepackage{tabularx}
\usepackage{booktabs}
\usepackage{multirow}
\usepackage{makecell}
\usepackage{amsmath} 

\usepackage{amssymb}  
\usepackage{url} 
\usepackage{microtype} 
\usepackage{placeins}
\usepackage{algorithm}
\usepackage{algpseudocode}
\usepackage{orcidlink}
\usepackage[
    colorlinks=true,
    linkcolor=blue,
    citecolor=blue,
    urlcolor=blue,
    bookmarks=true,
    bookmarksnumbered=true,
    pdfstartview=FitH
]{hyperref}
\usepackage{color}

\begin{document}
\title{Trans-RAG: Query-Centric Vector Transformation for
  Secure Cross-Organizational Retrieval}
\titlerunning{Trans-RAG: Query-Centric Vector Transformation}
%
%
\author{
Yu Liu$^{1,2}$\orcidlink{0009-0001-7592-8036} \and
Kun Peng$^{1,2}$\orcidlink{0000-0002-2051-4709} \and
Wenxiao Zhang$^{3}$\orcidlink{0009-0000-5196-8562} \and
Fangfang Yuan$^{1}$\textsuperscript{(\Letter)}\orcidlink{0000-0002-6368-8784} \and
Cong Cao$^{1}$\orcidlink{0000-0003-1881-1947} \and
Wenxuan Lu$^{1,2}$\orcidlink{0000-0002-8991-3978} \and
Yanbing Liu$^{1,2}$\textsuperscript{(\Letter)}\orcidlink{0000-0002-9653-073X}}

\authorrunning{Y. Liu et al.}

\institute{
$^{1}$Institute of Information Engineering, Chinese Academy of Sciences, Beijing, China\\
\email{\{liuyu,pengkun,yuanfangfang,caocong,luwenxuan,liuyanbing\}@iie.ac.cn}\\
$^{2}$School of Cyber Security, University of Chinese Academy of Sciences, Beijing, China\\
$^{3}$Department of Computer Science and Software Engineering, The University of Western Australia, Perth, Australia\\
\email{wenxiao.zhang@research.uwa.edu.au}}
\maketitle              
\begin{abstract}
Retrieval Augmented Generation (RAG) systems deployed across organizational boundaries face fundamental tensions between security, accuracy, and efficiency. Current Encryption methods expose plaintext during decryption, while federated architectures prevent resource integration and incur substantial overhead.
We introduce Trans-RAG, implementing a novel \textit{vector space language paradigm} where each organization's knowledge exists in a mathematically isolated semantic space. At the core lies vector2Trans, a multi-stage transformation technique that enables queries to dynamically "speak" each organization's vector space "language" through query-centric transformations, eliminating decryption overhead while maintaining native retrieval efficiency. Security evaluations demonstrate near-orthogonal vector spaces with 89.90° angular separation and 99.81\% isolation rates. Experiments across 8 retrievers, 3 datasets, and 3 LLMs show minimal accuracy degradation (3.5\% decrease in nDCG@10) and significant efficiency improvements over homomorphic encryption.
\keywords{Retrieval Augmented Generation \and Cross-organizational Security \and Vector Transformation \and Secure Information Retrieval \and Privacy-Preserving Embeddings}
\end{abstract}

\vspace{-35pt}
\begin{figure}[H]
  \centering
  \includegraphics[width=0.46\columnwidth]{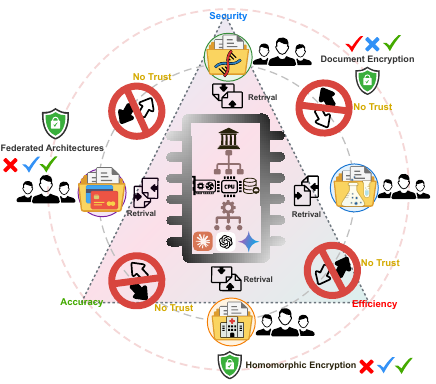}
  \vspace{-0.2cm}
  \caption{Cross-organizational retrieval with no trust faces the security-accuracy-efficiency triangle.}
  \label{fig:motivation}
  \vspace{-0.5cm}
\end{figure}

\section{Introduction}

Cross-organizational collaboration increasingly requires knowledge sharing 
while preserving data sovereignty across institutional boundaries. 
In healthcare, multi-institutional research must analyze distributed patient data 
without violating strict privacy regulations~\cite{brauneck2023federated,johner2025governance,kim2024federated}. 
In supply chains, partners integrate confidential supplier information 
while protecting competitive advantages~\cite{li2008confidentiality}. 

Recent advances in retrieval-augmented generation (RAG) further amplify 
the need for secure cross-organizational retrieval, as embedding-based systems 
introduce new privacy risks~\cite{lewis2020retrieval,zeng2024good}. 
As illustrated in Figure~\ref{fig:motivation}, cross-organizational retrieval 
without mutual trust inevitably encounters the security–accuracy–efficiency trade-off.

Current secure retrieval approaches face limitations across three critical dimensions. \textbf{1)} From a computational perspective, traditional encryption techniques necessitate decryption of multiple candidate documents during retrieval, introducing prohibitive overhead that recent studies show renders systems impractical for real-time applications~\cite{zhou2024piano,kim2022privacy}. \textbf{2)} From an architectural perspective, these document-centric security paradigms fundamentally misalign with the vector-centric nature of modern dense retrieval systems~\cite{karpukhin2020dense,wang2021milvus}, creating friction that manifests as performance bottlenecks and deployment complexities~\cite{sun2021practical}. \textbf{3)}From a security perspective, recent research demonstrates that vector embeddings themselves leak substantial semantic information~\cite{huang2024transferable,morris2023text}, with reconstruction attacks recovering significant portions of original text, revealing vulnerabilities that document-level protections cannot address~\cite{zhang2025universal}.

To address these challenges, we present Trans-RAG, a comprehensive framework implementing a vector space language paradigm that fundamentally shifts security from document-level protection to query-level transformation. Instead of encrypting stored documents, Trans-RAG creates organization-specific vector space "languages"—mathematically isolated semantic spaces where each organization's knowledge exists in a unique coordinate system, akin to natural languages providing security through mutual unintelligibility. At the core lies vector2Trans, a novel multi-stage transformation technique that enables queries to dynamically "speak" each organization's vector space language through combining orthogonal matrices, bounded non-linearity, key-based permutation, and cryptographic blinding. This query-centric approach aligns naturally with vector-based architectures~\cite{karpukhin2020dense}, eliminating traditional security-performance tensions while operating transparently with existing vector databases and embedding models. By maintaining separate transformation keys for each organization, Trans-RAG ensures complete organizational control over knowledge assets while enabling collaborative intelligence through secure query processing, requiring no modifications to organizational infrastructure. Our contributions are \textbf{three-fold}:
\begin{enumerate}[leftmargin=*, nosep]
\item We introduce Trans-RAG\footnote{Codes: https://github.com/Ameame1/TransRAG}, implementing query-centric vector transformations that shift security from document encryption to mathematical space isolation for cross-organizational retrieval.
\item We develop vector2Trans, a multi-stage, key-derived transformation technique combining orthogonal matrices, bounded non-linearity, key-based permutation, and cryptographic blinding to create computationally isolated vector spaces with minimal retrieval degradation.

\item Comprehensive theoretical analysis and experimental validation demonstrate strong security properties against reconstruction and probing attacks, minimal accuracy degradation across diverse retrievers and datasets, and substantial efficiency improvements over cryptographic alternatives, confirming deployment compatibility with existing infrastructure.
\end{enumerate}

\section{Related Work}

\subsection{RAG Systems and Limitations}
RAG has become a key paradigm for injecting external knowledge into LLMs~\cite{lewis2020retrieval}, but its assumption of open corpus access is at odds with cross-organizational data-sovereignty constraints. Prior work shows that well-crafted queries can elicit sensitive information~\cite{zeng2024good}. Existing defenses largely secure documents via access control or encryption; however, these document-centric mechanisms conflict with the vector-centric nature of dense retrieval~\cite{karpukhin2020dense,wang2021milvus}, leading to plaintext exposure during decryption and operational overhead. This gap motivates a security model aligned with vector operations rather than document plaintext.

\subsection{Secure Information Retrieval}
Secure retrieval spans trade-offs: access control and content encryption
expose plaintext during retrieval (or re-ranking), and even advanced
searchable encryption can leak sensitive signals~\cite{sun2021practical,gui2023rethinking}.
Homomorphic encryption offers stronger guarantees but remains impractical
for real-time vector retrieval due to computational overhead~\cite{zhou2024piano,kim2022privacy}. Other directions either depend on specialized
hardware (TEEs) or face efficiency bottlenecks in high-dimensional similarity
search (zero-knowledge systems)~\cite{li2021prism}.
Enterprise federated RAG typically focuses on result aggregation, not on
securing the vector retrieval process itself.

\subsection{Vector Transformations}
Vector transformations have been applied mainly to dimensionality reduction and model distillation~\cite{morris2023text}, with limited use as a security mechanism. Meanwhile, embeddings are vulnerable to reconstruction attacks that risk leakage of source content~\cite{huang2024transferable,zhang2025universal}. Although mathematical transformations have supported privacy in other domains---such as random projections for biometric template protection~\cite{ali2024cancelable} and orthogonal transforms for privacy-preserving data mining~\cite{liu2007random}---their application to forming isolated vector spaces for cross-organizational retrieval has seen limited exploration. 
\begin{figure*}[t]

  \centering
  \includegraphics[width=\textwidth]{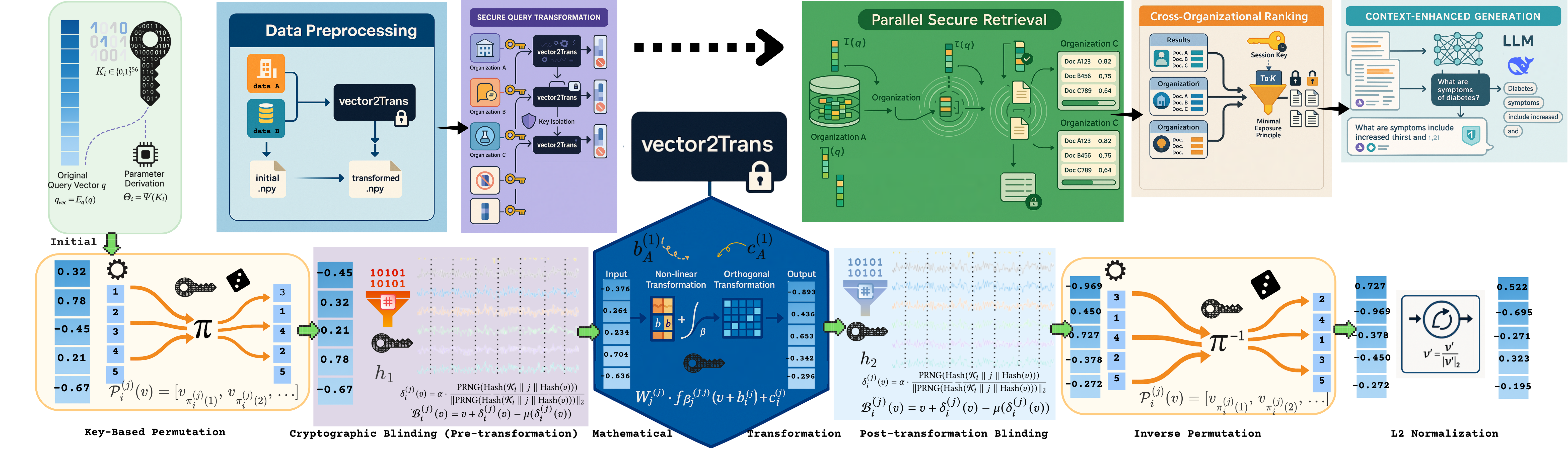}\
\caption{Trans-RAG overview and vector2Trans. \textbf{Top}: system workflow across organization-specific vector spaces (five phases from query encoding to context-enhanced generation) preserving data sovereignty. \textbf{Bottom}: \textit{vector2Trans}—a multi-stage, key-derived query transformation (permutation, cryptographic blinding, bounded non-linearity $f_\beta$, orthogonal rotation $W$, and L2 normalization) that yields computationally isolated spaces while retaining retrieval utility.}
  \label{fig:trans-rag-architecture}
\end{figure*}

\section{Methodology}

\subsection{Problem Formulation}

Given $m$ organizations $\{O_1,\ldots,O_m\}$, where each $O_i$ maintains a private vector database $V_i \subset \mathbb{R}^d$ built from embeddings of proprietary documents $D_i$, secure cross-organizational RAG aims to retrieve from the unified corpus $\bigcup_{i=1}^m V_i$ under two constraints: (i) no organization may access another's plaintext vectors; and (ii) retrieval accuracy is preserved within any space it is authorized to query. Traditional defenses encrypt documents or employ federated architectures, but these are misaligned with vector-centric dense retrieval and introduce operational overheads~\cite{karpukhin2020dense,wang2021milvus}. Trans-RAG addresses this by applying query-level transformations that yield per-organization, \emph{computationally isolated} vector spaces while retaining similarity structure for authorized retrieval.

\subsection{Threat Model and Security Objective}

\noindent\textbf{Adversary Model.} We adopt a semi-honest adversary model where: (1) organizations follow the protocol but may attempt to extract information from observed transformed vectors, (2) the adversary cannot access the transformation keys of other organizations, and (3) cryptographic primitives (e.g., SHA-256 and AES-based PRNGs) behave as pseudorandom functions.

\noindent\textbf{Security Objective.} The goal is \emph{computational isolation}: with distinct keys $\mathcal{K}_i \neq \mathcal{K}_j$, an adversary should gain negligible advantage in (a) reconstructing plaintext vectors, (b) inferring cross-space semantic relations, or (c) issuing effective cross-space queries without authorization.

\subsection{Trans-RAG Framework Overview}

As illustrated in Fig.~\ref{fig:trans-rag-architecture}, Trans-RAG operates through five phases: \textbf{(1) Query Encoding}---the user query is encoded into a dense vector $v_q \in \mathbb{R}^d$ using an embedding model/encoder; \textbf{(2) Query Transformation}---$v_q$ is transformed into organization-specific representations $\{\mathcal{T}_i(v_q)\}$ using each organization's key $\mathcal{K}_i$; \textbf{(3) Parallel Retrieval}---transformed queries are dispatched to authorized organizations' vector stores in parallel, returning top-$k$ candidates with similarity scores; \textbf{(4) Context Aggregation}---scores are normalized across spaces, results are re-ranked, and access control is verified; \textbf{(5) Answer Generation}---aggregated contexts are forwarded to the LLM for final response synthesis. The core security mechanism lies in the transformation $\mathcal{T}_i(\cdot)$, detailed below.

\subsection{vector2Trans: Query-Centric Transformation}

\noindent\textbf{Overview and design principles.}\,Vector2Trans creates per-organization isolated spaces via multi-stage, key-derived transformations. As illustrated in Fig.~\ref{fig:trans-rag-architecture}, each stage applies, in order: key-based permutation (decorrelate coordinates), input-dependent cryptographic blinding (disrupt statistics), bounded non-linearity (similarity-preserving yet non-invertible), and orthogonal rotation with offsets (structure-preserving mixing). All parameters derive deterministically from the organization key $\mathcal{K}_i \in \{0,1\}^{256}$.

\paragraph{Mathematical formulation.}
The $n$-stage transformation is
\begin{equation}
\label{eq:transform_composite}
\mathcal{T}_i(v;\mathcal{K}_i) = \mathcal{T}_i^{(n)} \circ \mathcal{T}_i^{(n-1)} \circ \cdots \circ \mathcal{T}_i^{(1)}(v).
\end{equation}
Each stage $j \in \{1,\ldots,n\}$ integrates key-based permutation $\mathcal{P}_i^{(j)}$, cryptographic blinding $\mathcal{B}_i^{(j)}$, orthogonal rotation $W_i^{(j)}$, bounded non-linearity $f_{\beta_i^{(j)}}$, and offset vectors $b_i^{(j)}, c_i^{(j)}$:
\begin{equation}
\label{eq:transform_stage}
\mathcal{T}_i^{(j)}(v) = \mathcal{P}_i^{(j)-1}(\mathcal{B}_i^{(j+n)}(W_i^{(j)} \cdot f_{\beta_i^{(j)}}(\mathcal{B}_i^{(j)}(\mathcal{P}_i^{(j)}(v)) + b_i^{(j)}) + c_i^{(j)})),
\end{equation}
followed by $\ell_2$-normalization.

\noindent\textbf{Component definitions.} The permutation operator $\mathcal{P}_i^{(j)}: \mathbb{R}^d \to \mathbb{R}^d$ reorders vector dimensions:
\begin{equation}
\mathcal{P}_i^{(j)}(v) = [v_{\pi_i^{(j)}(1)}, v_{\pi_i^{(j)}(2)}, \ldots, v_{\pi_i^{(j)}(d)}],
\end{equation}
where $\pi_i^{(j)}: \{1, \ldots, d\} \to \{1, \ldots, d\}$ is a cryptographically derived permutation: $\pi_i^{(j)} = \text{Shuffle}(\text{PRNG}(\text{Hash}(\mathcal{K}_i \parallel j)))$, and $\mathcal{P}_i^{(j)-1}$ denotes its inverse.

The cryptographic blinding operator $\mathcal{B}_i^{(j)}: \mathbb{R}^d \to \mathbb{R}^d$ adds zero-mean input-dependent noise:
\begin{equation}
\mathcal{B}_i^{(j)}(v) = v + \delta_i^{(j)}(v) - \mu(\delta_i^{(j)}(v)),
\end{equation}
where $\mu(\cdot)$ denotes the mean (ensuring zero-mean noise), and the blinding noise is generated as:
\begin{equation}
\delta_i^{(j)}(v) = \alpha \cdot \frac{\text{PRNG}(\text{Hash}(\mathcal{K}_i \parallel j \parallel \text{Hash}(v)))}{\|\text{PRNG}(\text{Hash}(\mathcal{K}_i \parallel j \parallel \text{Hash}(v)))\|_2},
\end{equation}
with $\alpha > 0$ controlling the blinding intensity.

The mathematical transformation combines an orthogonal matrix $W_i^{(j)} \in \mathbb{R}^{d \times d}$ satisfying $(W_i^{(j)})^{\top}W_i^{(j)} = I$, generated via QR decomposition of pseudorandom matrices seeded by $\text{Hash}(\mathcal{K}_i \parallel j)$; a bounded non-linear function $f_{\beta}(x) = \tanh(\beta x)/\beta$ where $\beta > 0$ is the non-linearity strength parameter; and offset vectors $b_i^{(j)}, c_i^{(j)} \in \mathbb{R}^d$ derived from the organization key. The final output of all $n$ stages undergoes $\ell_2$-normalization to ensure unit length.

\subsection{Security Properties}

\begin{property}[Cross-Space Angular Separation]\label{prop:angular_separation}
For distinct keys $\mathcal{K}_i \neq \mathcal{K}_j$ and unit-norm vectors $v$ drawn from a bounded distribution,
\[
\left|\mathbb{E}\big[\cos\!\big(\mathcal{T}_i(v;\mathcal{K}_i),\mathcal{T}_j(v;\mathcal{K}_j)\big)\big]\right|=O(d^{-1/2}).
\]
\end{property}
\begin{proof}[Sketch]
Under the PRF assumption (independent key-derived permutations and orthogonals), the outputs are independent random rotations plus zero-mean input-dependent blindings; the bounded odd non-linearity $f_\beta$ is $1$-Lipschitz. By concentration on $\mathbb{S}^{d-1}$, for $x,y\!\sim\!\mathrm{Unif}(\mathbb{S}^{d-1})$ we have $\mathbb{E}\langle x,y\rangle=0$ and $\mathrm{Var}(\langle x,y\rangle)=1/d$~\cite{vershynin2018high}. Hence $\big|\mathbb{E}[\cos]\big|=O(d^{-1/2})$.
\end{proof}

\begin{property}[Reconstruction Resistance]\label{prop:reconstruction}
Without $\mathcal{K}_i$, reconstructing $v$ from $\mathcal{T}_i(v)$ requires solving a mixed discrete--continuous, non-convex inverse problem with input-dependent noise; any efficient inversion would contradict standard PRF assumptions and is therefore computationally infeasible.
\end{property}
\begin{proof}[Sketch]
Per stage, the Jacobian is $J^{(j)}(v)=\mathcal{P}^{(j)-1} D^{(j)}(v) W^{(j)} \mathcal{P}^{(j)}$ (up to affine shifts/blindings), where $D^{(j)}(v)=\mathrm{diag}(f'_{\beta^{(j)}}(\cdot))$ and $f'_\beta(z)=\mathrm{sech}^2(\beta z)\!\in\!(0,1]$. Thus $\|J^{(j)}\|_2\!\le\!1$ and saturations yield information loss; over $n$ stages the composition is further contracting on saturated coordinates (non-invertible). Inversion must simultaneously recover unknown permutations (discrete, exponential), orthogonals on $O(d)$ (continuous), and input-dependent blindings tied to $\mathrm{Hash}(v)$ (PRF). Any efficient inverter would distinguish PRF outputs from random, contradicting the assumption.
\end{proof}

\begin{property}[Cross-Organizational Query Isolation]\label{prop:query_isolation}
A query transformed with $\mathcal{K}_i$ yields near-random similarity scores in $V_j$ ($j\!\neq\! i$), producing random-like retrieval.
\end{property}
\begin{proof}[Sketch]
From Prop.~\ref{prop:angular_separation}, for $\hat q=\mathcal{T}_i(q)$ and $\hat v=\mathcal{T}_j(v)$ we have $\langle \hat q,\hat v\rangle$ concentrated near $0$ with variance $O(1/d)$. Hence scores are sub-Gaussian; the maximum over $N$ items scales as $\tilde O(\sqrt{\log N/d})$, matching random retrieval and strictly below authorized-space signal where aligned transforms preserve neighbors.
\end{proof}

\smallskip
\noindent\textbf{Scope \& assumptions.}
These arguments rely on: (i) PRF-derived independence across keys; (ii) high-dimensional concentration; (iii) bounded, odd $f_\beta$ and zero-mean blindings; and (iv) semi-honest parties not influencing PRF seeds.
Worst-case/adaptive attacks that manipulate input distributions or observe longitudinal correlations are out of scope here and evaluated empirically in Sec.~\ref{sec:practical_attacks}.

\section{Experiments}
\subsection{Experimental Setup}

\noindent\textbf{Datasets and Settings.}
We evaluate retrieval effectiveness using three BEIR datasets~\cite{thakur2021beir}: \textbf{NFCorpus}~\cite{boteva2016nfcorpus} (3{,}633 docs; 323 queries), \textbf{FiQA}~\cite{maia2018fiqa} (5{,}778 docs; 648 queries), and \textbf{SciFact}~\cite{wadden2020scifact} (5{,}183 docs; 1{,}109 queries). In the cross-organizational setting ($m\!=\!10$), each dataset is partitioned across organizations via stratified sampling (preserving label/topic distributions); unless otherwise specified, all organizations are authorized for querying. For efficiency and scaling evaluation, we construct auxiliary corpora with sizes \{1K, 10K, 100K, 1000K\} documents, used exclusively for latency and throughput tests (disjoint from BEIR).

\noindent\textbf{Models.}
We evaluate 8 dense retrievers across different embedding dimensions: Ember, UAE, GTE~\cite{li2023gte} (1024d), MPNet~\cite{song2020mpnet}, BGE~\cite{chen2024bge} (768d), Jasper~\cite{zhang2024jasper} (512d), Linq~\cite{choi2024linq} (4096d), and Stella~\cite{zhang2024jasper} (8192d). For answer synthesis, we use generation models LLaMA 3.1-8B~\cite{dubey2024llama}, DeepSeek-V3~\cite{deepseek2024v3}, and Claude Sonnet 4~\cite{anthropic2025claude4}. Query-side transformation latency is independent of the LLM.

\noindent\textbf{Baselines and Metrics.}
We compare Trans-RAG against:
(1) \textbf{Naive Retrieval} (no security) as an upper bound~\cite{karpukhin2020dense};
(2) \textbf{AEAD} (AES-GCM-256) content protection~\cite{sun2021practical}, where ranking is performed on decrypted embeddings (ranking accuracy is unchanged);
(3) \textbf{Partial Homomorphic Encryption (PHE)} using Paillier-2048~\cite{paillier1999public} for encrypted dot-product similarity (Python phe library).
We evaluate retrieval effectiveness using nDCG@10, top-$k$ overlap with the naive baseline (Jaccard), and Spearman correlation~\cite{spearman1904proof}.
Efficiency metrics include per-query transformation and retrieval latency, scaling behavior with corpus size, and incremental update time.
For security, we report cross-space angular separation, isolation rate, $k$-NN neighborhood purity/preservation, and information-theoretic measures (entropy, mutual information, KL divergence).

\noindent\textbf{Implementation Details.}
For similarity computation, we use cosine similarity with $\ell_2$-normalized embeddings and top-$k\!=\!10$ results.
Trans-RAG employs a 256-bit key seed, with $n\!=\!3$ transformation stages, non-linearity $\beta\!=\!0.1$, and blinding intensity $\alpha\!=\!0.1$; retriever dimensions are matched to $d$.
We use FAISS~\cite{johnson2019billion,douze2024faiss} for indexing: IndexFlatIP for BEIR datasets, and IndexIVFFlat with $nlist=100$ and $nprobe=10$ for scalability experiments.
All experiments are repeated 10 times with fixed random seeds, with standard deviations of less than 2\% across all metrics. Mutual information is estimated using a k-NN entropy estimator averaged across dimensions.
We used 4$\times$NVIDIA A100 for indexing/embedding; CPU specs for PHE/AEAD are Intel Xeon Gold 6248R with 128 GB RAM.
\vspace{-0.4cm}
\subsection{Retrieval Accuracy}

\begin{table*}[t]
  \centering
  \footnotesize
  \caption{Retrieval accuracy across datasets (nDCG@10/Overlap\%/Spearman)}
  \label{tab:retrieval-acc}
  \resizebox{\textwidth}{!}{%
    \begin{tabular}{ll|cccccccc|c}
      \toprule
      \textbf{Dataset} & \textbf{Method} & \textbf{Ember} & \textbf{UAE} & \textbf{GTE} & \textbf{MPNet} & \textbf{BGE} & \textbf{Jasper} & \textbf{Stella} & \textbf{Linq} & \textbf{Avg} \\
      & & \textbf{(1024d)} & \textbf{(1024d)} & \textbf{(1024d)} & \textbf{(768d)} & \textbf{(768d)} & \textbf{(512d)} & \textbf{(8192d)} & \textbf{(4096d)} & \\
      \midrule
      \multirow{2}{*}{\textbf{NFCorpus}}
      & Naive & 0.314/100.0/1.000 & 0.320/100.0/1.000 & 0.204/100.0/1.000 & 0.289/100.0/1.000 & 0.362/100.0/1.000 & 0.390/100.0/1.000 & 0.227/100.0/1.000 & 0.310/100.0/1.000 & 0.302/100.0/1.000 \\
      & Trans-RAG & 0.303/92.8/0.860 & 0.309/92.7/0.845 & 0.197/92.8/0.855 & 0.279/92.4/0.811 & 0.349/91.8/0.762 & 0.375/92.3/0.804 & 0.219/93.7/0.931 & 0.299/92.9/0.862 & \textbf{0.291/92.7/0.841} \\
      \midrule
      \multirow{2}{*}{\textbf{FiQA}}
      & Naive & 0.443/100.0/1.000 & 0.445/100.0/1.000 & 0.228/100.0/1.000 & 0.496/100.0/1.000 & 0.424/100.0/1.000 & 0.556/100.0/1.000 & 0.200/100.0/1.000 & 0.460/100.0/1.000 & 0.407/100.0/1.000 \\
      & Trans-RAG & 0.427/92.5/0.856 & 0.430/92.4/0.847 & 0.220/92.3/0.843 & 0.478/92.4/0.851 & 0.408/92.4/0.850 & 0.535/91.2/0.737 & 0.194/93.2/0.921 & 0.444/92.7/0.858 & \textbf{0.392/92.4/0.845} \\
      \midrule
      \multirow{2}{*}{\textbf{SciFact}}
      & Naive & 0.737/100.0/1.000 & 0.743/100.0/1.000 & 0.354/100.0/1.000 & 0.634/100.0/1.000 & 0.747/100.0/1.000 & 0.770/100.0/1.000 & 0.531/100.0/1.000 & 0.725/100.0/1.000 & 0.655/100.0/1.000 \\
      & Trans-RAG & 0.712/93.2/0.844 & 0.716/93.2/0.843 & 0.343/93.4/0.857 & 0.613/92.9/0.809 & 0.721/92.7/0.791 & 0.745/92.2/0.747 & 0.513/94.4/0.950 & 0.701/93.2/0.841 & \textbf{0.633/93.2/0.835} \\
      \midrule
      \multirow{2}{*}{\textbf{Overall}}
      & Naive & 0.498/100.0/1.000 & 0.503/100.0/1.000 & 0.262/100.0/1.000 & 0.473/100.0/1.000 & 0.511/100.0/1.000 & 0.572/100.0/1.000 & 0.319/100.0/1.000 & 0.498/100.0/1.000 & 0.455/100.0/1.000 \\
      & Trans-RAG & 0.481/92.8/0.853 & 0.485/92.8/0.845 & 0.253/92.8/0.852 & 0.457/92.6/0.824 & 0.493/92.3/0.801 & 0.552/91.9/0.763 & 0.309/93.8/0.934 & 0.481/92.9/0.854 & \textbf{0.439/92.8/0.840} \\
      \midrule
      \multirow{1}{*}{\textbf{Reduction}}
      & Trans-RAG & ↓3.4\%/↓7.2\%/↓14.7\% & ↓3.6\%/↓7.2\%/↓15.5\% & ↓3.4\%/↓7.2\%/↓14.8\% & ↓3.4\%/↓7.4\%/↓17.6\% & ↓3.5\%/↓7.7\%/↓19.9\% & ↓3.5\%/↓8.1\%/↓23.7\% & ↓3.1\%/↓6.2\%/↓6.6\% & ↓3.4\%/↓7.1\%/↓14.6\% & \textbf{↓3.5\%/↓7.2\%/↓16.0\%} \\
      \midrule
      \multicolumn{10}{l}{$^\ast \text{Naive}_{\text{nDCG@10/Overlap/Spearman}}\approx \text{PHE}_{\text{nDCG@10/Overlap/Spearman}}\approx\text{AEAD}_{\text{nDCG@10/Overlap/Spearman}}$.}
    \end{tabular}
  }
  \vspace{-0.6cm}

\end{table*}
We evaluate Trans-RAG's retrieval quality across multiple datasets and retrievers, comparing it with unprotected and alternative secure methods. As shown in Table~\ref{tab:retrieval-acc}, Trans-RAG effectively preserves retrieval quality, with nDCG@10 scores showing minimal degradation of only 3.1\%--3.6\% (average 3.5\%) compared to the unprotected Naive baseline. The absolute differences range from 0.011 to 0.022, confirming that Trans-RAG maintains strong retrieval performance while providing robust security guarantees. Additionally, overlap coefficients range from 92.4\% to 93.2\%, representing reductions of 6.2\% to 8.1\% (average 7.2\%), and Spearman correlation coefficients range from 0.835 to 0.845, showing reductions of 6.6\% to 23.7\% (average 16.0\%) across all datasets.

High-dimensional retrievers show exceptional compatibility with Trans-RAG---particularly Stella (8192d), which achieves overlap coefficients as high as 94.4\% and Spearman coefficients up to 0.950. This superior performance with high-dimensional embeddings can be attributed to greater redundancy in high-dimensional spaces, allowing better preservation of semantic structure after transformation.

Our analysis reveals consistently strong performance across diverse retrieval scenarios, with minimal variation across configurations (overlap coefficient standard deviation <0.6\%). Trans-RAG's model-agnostic design ensures compatibility with embeddings ranging from 512d to 8192d, making it a versatile security solution adaptable to various retrieval architectures.
\vspace{-0.3cm}
\subsection{Security Evaluations}

We rigorously evaluated Trans-RAG's security properties across multiple dimensions, focusing on vector space separation, neighborhood structure, and information-theoretic properties.

\noindent \textbf{Vector Space Isolation.}
We measure pairwise angular separation across 10 organizations before and after transformation.

\begin{figure}[t]
  \centering
  \includegraphics[width=1\columnwidth]{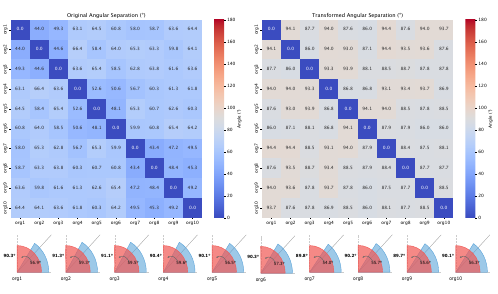}
  \caption{Angular separation between vector spaces before and after transformation. The transformation increases average separation from 58.33° to 89.90°, approaching perfect orthogonality. Avg Cosine Similarity: 0.506→0.009 (Impr: 0.497).}
  \label{fig:angle_separation_heatmap}
\end{figure}

Fig.~\ref{fig:angle_separation_heatmap} shows the transformation increases angular separation from 58.33° to 89.90°, achieving near-orthogonal vector spaces across all organization pairs.

\begin{figure}[H]
\centering
\includegraphics[width=1\columnwidth]{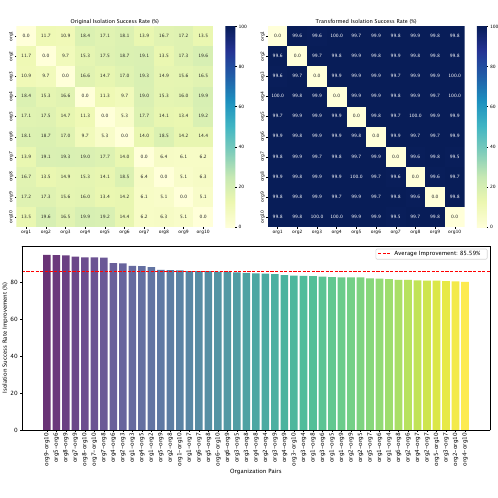}
\caption{Isolation success rates (Cosine Similarity <0.1) across organization pairs. All pairs achieve isolation rates exceeding 99.5\% (Minimum) after transformation. Avg: 14.22\%→99.81\% (Impr: 85.59\%).}
\label{fig:isolation_heatmap}
\end{figure}

To quantify this isolation more precisely, we compute the isolation success rate---the proportion of vector pairs with cosine similarity below 0.1. Fig.~\ref{fig:isolation_heatmap} shows that all organization pairs achieve isolation rates exceeding 99.5\% after transformation.

These results empirically validate Properties~\ref{prop:angular_separation} and~\ref{prop:query_isolation}. The 89.90° angular separation closely aligns with the theoretical $O(d^{-1/2})$ expected cosine similarity for high-dimensional vectors, confirming near-orthogonal isolation.

\noindent \textbf{Neighborhood Structure Analysis.} Beyond global separation, we analyze local topology via $k$-nearest neighbors ($k$-NN). Specifically, we compute: 

(1) \textit{neighborhood purity} as the percentage of same-organization neighbors, defined as $\mathrm{Purity}_i = \frac{1}{N_i}\sum_{p=1}^{N_i}\frac{|\{j\in \mathrm{NN}[p]: j\in \mathrm{org}_i\}|}{k}$.

(2) \textit{neighborhood preservation} as the overlap between original and transformed neighbors, given by $\mathrm{Preservation}_i = \frac{1}{N_i}\sum_{p=1}^{N_i}\frac{|\mathrm{NN}^{\mathrm{orig}}
[p]\cap\mathrm{NN}^{\mathrm{trans}}[p]|}{k}$.

(3) \textit{disturbance rate}, quantifying topological change.

\begin{table}[t]
  \centering
  \caption{Neighborhood structure metrics before and after transformation}
  \label{tab:neighbor_distribution}
  \resizebox{\columnwidth}{!}{%
    \begin{tabular}{lcccccccccc|c}
      \hline
      \textbf{Metric} & \textbf{org1} & \textbf{org2} & \textbf{org3} & \textbf{org4} & \textbf{org5} & \textbf{org6} & \textbf{org7} & \textbf{org8} & \textbf{org9} & \textbf{org10} & \textbf{Avg} \\
      \hline
      Original Purity (\%)           & 43.8  & 26.2  & 25.1  & 34.6  & 42.1  & 44.8  & 33.6  & 37.2  & 32.5  & 32.5  & 35.24 \\
      Transformed Purity (\%)        & 100.0 & 100.0 & 100.0 & 100.0 & 100.0 & 100.0 & 100.0 & 100.0 & 100.0 & 100.0 & 100.0 \\
      Neighborhood Preservation (\%) & 22.51 & 25.60 & 32.92 & 22.74 & 26.20 & 38.27 & 28.63 & 24.26 & 28.56 & 30.29 & 28.00 \\
      Disturbance Rate (\%)          & 56.2  & 73.8  & 74.9  & 65.4  & 57.9  & 55.2  & 66.4  & 62.8  & 67.5  & 67.5  & 64.76 \\
      \hline
      \multicolumn{12}{l}{$^{*}$ $\displaystyle \mathrm{Disturbance}_i = \mathrm{Purity}_i^{\mathrm{trans}} - \mathrm{Purity}_i^{\mathrm{orig}}$.}
    \end{tabular}%
  }
\end{table}

\begin{figure}[t]
  \centering
  \includegraphics[width=1\columnwidth]{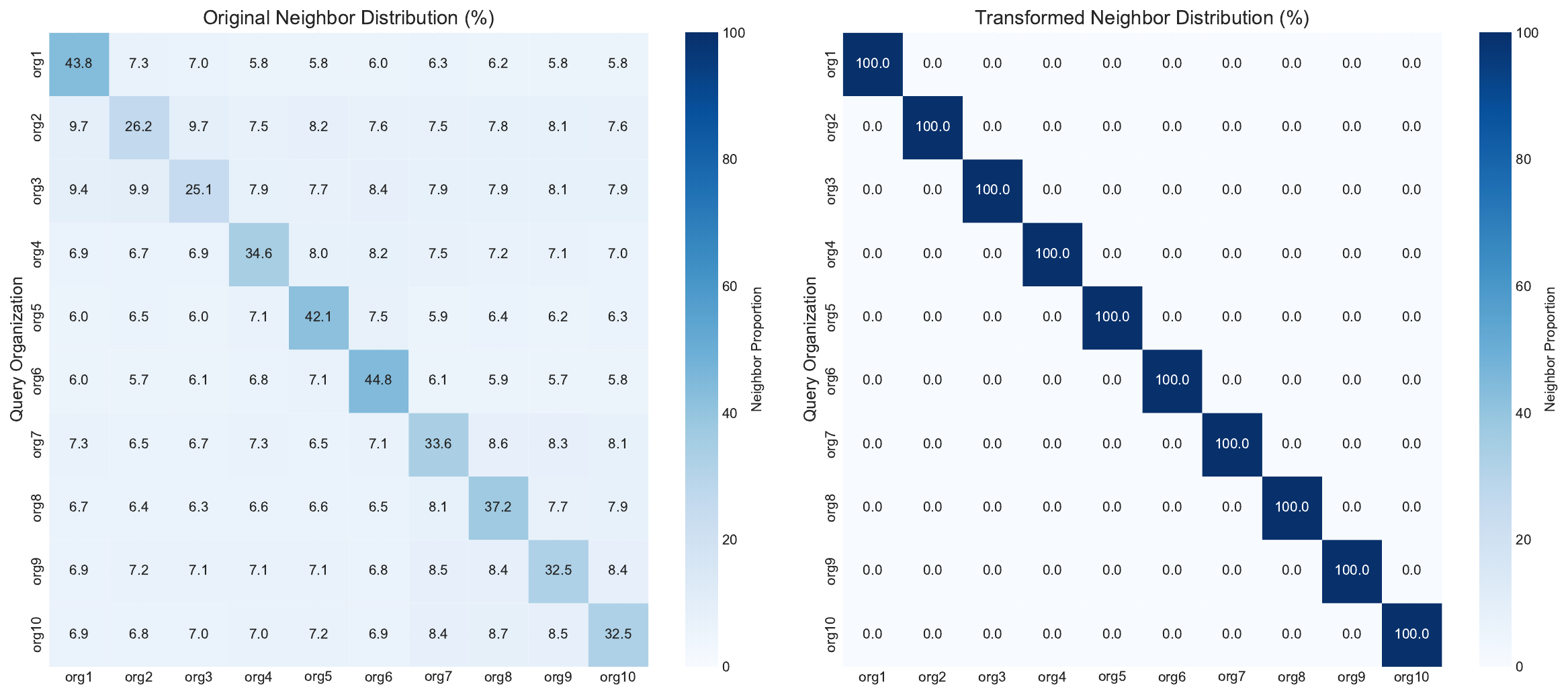}
  \caption{Neighborhood purity before and after transformation. Avg (35.24\%→100\%)}
  \label{fig:neighbor_distribution}

\end{figure}

Fig.~\ref{fig:neighbor_distribution} reveals the transformation's impact on local vector relationships using $k=20$. The original space shows substantial cross-organizational mixing (diagonal elements: 25.1\%--44.8\%), while the transformed space achieves perfect diagonal isolation (100\%), eliminating cross-organizational nearest neighbors.

Table~\ref{tab:neighbor_distribution} confirms Property~\ref{prop:query_isolation} with 100\% neighborhood purity and 28.00\% preservation rate. The disturbance rate (avg. 64.76\%) varies adaptively, balancing organizational separation with intra-organizational retrieval utility.

\noindent \textbf{Information-Theoretic Analysis.}
Information theory provides a rigorous framework for quantifying vector2Trans's security-utility balance~\cite{Shannon1948,Cover2006}. We analyze entropy, self-mutual information~\cite{Kong2022}, and Kullback-Leibler divergence~\cite{Kullback1951}.

Self-mutual information (SMI) measures shared information between original and transformed spaces, and is defined as $\mathrm{MI}(V_i^{\mathrm{orig}},V_i^{\mathrm{trans}}) = \frac{\sum_{j=1}^n \mathrm{MI}_j \lambda_j}{\sum_{j=1}^n \lambda_j}$, where $\mathrm{MI}_j$ represents the mutual information of the $j$-th component, and $\lambda_j$ are the corresponding weights.

Near-identical entropy values (9.909 vs. 9.911 bits) confirm information capacity preservation (Table~\ref{tab:info_theory}). Self-mutual information of 6.141 bits (62.0\% preservation) enables effective retrieval while concealing sufficient information for security. High KL divergence (24.960 bits) establishes a statistical barrier between distributions, quantifying the security-utility balance.

\begin{table}[t]
  \centering
  \caption{Information-theoretic measurements across organizations}
  \label{tab:info_theory}
  \resizebox{\columnwidth}{!}{%
    \begin{tabular}{l|cccccccccc|c}
      \hline
      \textbf{Metric} & \textbf{org1} & \textbf{org2} & \textbf{org3} & \textbf{org4} & \textbf{org5} & \textbf{org6} & \textbf{org7} & \textbf{org8} & \textbf{org9} & \textbf{org10} & \textbf{Avg} \\
      \hline
      Entropy-Orig (bits) & 9.912 & 9.917 & 9.917 & 9.904 & 9.880 & 9.913 & 9.952 & 9.866 & 9.946 & 9.879 & 9.909 \\
      Entropy-Trans (bits) & 9.915 & 9.920 & 9.919 & 9.906 & 9.882 & 9.915 & 9.954 & 9.869 & 9.948 & 9.881 & 9.911 \\
      Self-MI (bits) & 6.226 & 6.265 & 6.146 & 6.280 & 6.148 & 6.121 & 6.171 & 6.048 & 6.015 & 5.985 & 6.141 \\
      KL Divergence (bits) & 24.544 & 24.243 & 26.598 & 25.592 & 23.868 & 26.251 & 24.081 & 24.394 & 24.720 & 25.306 & 24.960 \\
      \hline
    \end{tabular}%
  }

\end{table}

\begin{table}[t]
  \centering
  \small  
  \caption{Efficiency vs. scale by retriever (ms)}
  \label{tab:efficiency_comparison}
  \setlength{\tabcolsep}{4.5pt}
  \begin{tabular*}{\linewidth}{@{\extracolsep{\fill}} lccc|ccc}
    \toprule
    \multirow{2}{*}{Method} & \multicolumn{3}{c|}{Query Latency Overhead} & \multicolumn{3}{c}{Data Processing Overhead} \\
    \cmidrule(lr){2-4} \cmidrule(l){5-7}
     & 10K & 100K & 1000K & 10K & 100K & 1000K \\
    \midrule
    Trans-RAG (Avg)   & 1,117.5 & 1,007.5 & 999.5  & 113.4 & 335.5 & 1,037.3 \\
    PHE (Avg)         & 3.6$\times$10$^{7}$ & NA & NA & 4.4$\times$10$^{7}$ & NA & NA \\
    AEAD (Avg)        & 286.4 & 3,127.6 & 38,752.8 & 68.6 & 203.4 & 626.8 \\
    \midrule
    Trans-RAG (768d)  & 73.3 & 68.7 & 70.7 & 72.9 & 215.6 & 665.8 \\
    Trans-RAG (1024d) & 111.6 & 103.6 & 95.9 & 105.3 & 311.6 & 962.1 \\
    Trans-RAG (4096d) & 1,635.4 & 1,504.6 & 1,482.0 & 148.7 & 439.7 & 1,357.9 \\
    Trans-RAG (8192d) & 6,649.9 & 6,016.7 & 6,164.8 & 135.5 & 400.7 & 1,237.3 \\
    \bottomrule
  \end{tabular*}

  {\raggedright\footnotesize
  $^\ast$ $\mathrm{Time}_{\text{overhead}}=\mathrm{Time}_{\text{baseline}}-\mathrm{Time}_{\text{naive}}$.\par}

\end{table}

\begin{table*}[t]
  \centering
  \small  
  \caption{Trans-RAG query latency overhead across retrievers and LLMs (ms)}
  \label{tab:retriever_llm_performance}
  \setlength{\tabcolsep}{4pt}
  \begin{tabular*}{\textwidth}{@{\extracolsep{\fill}} l|cccccccc}
    \toprule
    Models & Ember & UAE & GTE & MPNet & BGE & Jasper & Stella & Linq \\
    \midrule
    LLaMA 3.1-8B    & 112.3 & 108.7 & 103.2 & 71.2 & 70.5 & 58.7 & 6,235.4 & 1,527.3 \\
    DeepSeek-V3 & 109.5 & 103.6 & 106.8 & 68.7 & 73.8 & 55.5 & 6,016.7 & 1,504.6 \\
    Claude Sonnet 4  & 113.8 & 105.9 & 101.4 & 72.8 & 69.3 & 57.1 & 6,329.2 & 1,513.8 \\
    \bottomrule
  \end{tabular*}

  \vspace{2pt}
  {\raggedright\footnotesize
  $^\ast$ $\mathrm{Time}_{\text{overhead}}=\mathrm{Time}_{\text{baseline}}-\mathrm{Time}_{\text{naive}}$.\par}

\end{table*}

\begin{table}[t]
  \centering
  \footnotesize
  \caption{Document addition processing time comparison (in seconds)}
  \label{tab:processing_time_comparison}
  \resizebox{\columnwidth}{!}{%
    \begin{tabular}{lcc|c}
      \hline
      Scenario (Original + New Docs) & Full Rebuild & Incremental Update & Speedup \\
      \hline
      10K + 1K (11K total) & 228.7 & 36.2 & 6.3× \\
      100K + 1K (101K total) & 2,314.4 & 36.2 & 63.9× \\
      100K + 10K (110K total) & 2,517.5 & 363.2 & 6.9× \\
      1000K + 1K (1001K total) & 27,509.0 & 36.2 & 760.0× \\
      1000K + 10K (1010K total) & 27,623.1 & 362.0 & 76.3× \\
      1000K + 100K (1100K total) & 28,765.2 & 2,671.1 & 10.8× \\
      \hline
      \multicolumn{4}{l}{$^\ast$ Results averaged across retrievers from 512d to 8192d dimensions.} \\
      \multicolumn{4}{l}{Values are rounded; identical entries may reflect different underlying measurements.}
    \end{tabular}%
  }
  \vspace{-0.5cm}
\end{table}

\noindent \textbf{Scalability to Large Consortia.}
Although our experiments involve 10 organizations,
Property~\ref{prop:angular_separation} indicates that the expected
cross-space cosine similarity decays as $O(d^{-1/2})$ and remains
independent of the organization count $m$ under key independence
assumptions. Thus, isolation quality is governed primarily by
embedding dimensionality rather than consortium size.

Under parallel execution, query dispatch incurs $O(m)$ communication
cost, while result aggregation scales as $O(mk)$.
Since $k$ is constant and typically much smaller than $|V_i|$ in practical retrieval settings,
the overall overhead grows approximately linearly with $m$,
supporting practical deployment for large-scale consortia.

\subsection{Retrieval Efficiency}
Table~\ref{tab:efficiency_comparison} reveals distinct performance profiles: Trans-RAG achieves balanced latency (1,117.5 ms at 10K docs, 999.5 ms at 1000K) with minimal overhead for standard dimensions (68.7--111.6 ms for 768d--1024d). Table~\ref{tab:retriever_llm_performance} confirms consistency across LLM architectures (55.5--113.8 ms overhead for LLaMA 3.1-8B, DeepSeek-V3, Claude Sonnet 4). While standard dimensions achieve sub-100ms overhead, ultra-high-dimensional retrievers like Stella (8192d) exhibit ~6s latency due to the $O(nd^2)$ cost of multi-stage orthogonal transformations. Mitigation strategies such as GPU optimization or dimensionality reduction remain promising future directions.

PHE introduces prohibitive computation (3.6×10$^7$ ms at 10K docs, 32,216× slower than Trans-RAG), rendering it impractical despite formal cryptographic guarantees. AEAD shows low initial latency (286.4 ms at 10K) but scales poorly (38,752.8 ms at 1000K) and requires decrypting all candidates, exposing 100\% plaintext across organizational boundaries.

\noindent \textbf{Overhead Source Analysis.}
Overhead sources differ fundamentally: \textbf{Trans-RAG} incurs $O(nd^2)$ transformation cost plus TopK aggregation. \textbf{PHE} requires homomorphic dot products with $O(d \cdot k^3)$ complexity over 2048-bit integers, prohibitive for high-dimensional vectors. \textbf{AEAD} decrypts all candidates, with overhead scaling linearly with document count.

\noindent \textbf{Data Processing Efficiency.}
Table~\ref{tab:processing_time_comparison} shows incremental updates achieve 760.0× speedup when adding 1K documents to 1000K repositories, beneficial for production environments with frequent updates.

\subsection{Security Against Practical Attacks}
\label{sec:practical_attacks}

We evaluate vector2Trans under the semi-honest adversary model (Sec.~3.2) against cross-organizational probing, which directly tests Property~\ref{prop:query_isolation}. Experiments employ 10 organizations (90 directed pairs) with 100K documents per organization.

\subsubsection{Cross-organizational probing attack analysis}
\noindent \textbf{Setup.} For each directed pair $(O_i\rightarrow O_j)$ with $i\neq j$, we transform queries using $\mathcal{K}_i$ and retrieve from $V_j$.

\noindent \textbf{Results.}
\begin{figure}[t]
  \centering
  \includegraphics[width=1\columnwidth]{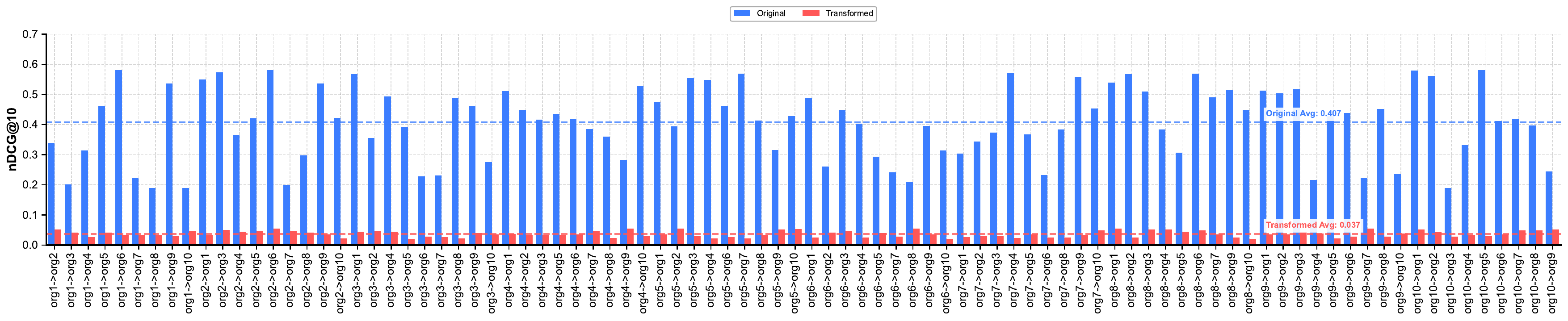}
  \vspace{-0.2cm}
  \caption{Cross-organizational probing attack analysis (10 organizations, 90 directed pairs).}
  \label{fig:probing_attack}
  \vspace{-0.4cm}
\end{figure}

\begin{table}[t]
\centering
\footnotesize
\caption{Ablation Study of vector2Trans Components (Accuracy \& Efficiency)}
\label{tab:ablation_study}
\resizebox{\textwidth}{!}{%
\begin{tabular}{l|ccc|cc}
\hline
\multirow{2}{*}{\textbf{Model Configuration}}
  & \multicolumn{3}{c|}{\textbf{Retrieval Accuracy}}
  & \multicolumn{2}{c}{\textbf{Computational Efficiency}} \\
\cline{2-6}
& \textbf{nDCG@10} & \textbf{Overlap \%} & \textbf{Spearman}
  & \textbf{Query (ms)} & \textbf{Process (ms)} \\
\hline
\textbf{Complete vector2Trans}
  & 0.409 & 92.2\% & 0.829
  & 108.6 & 109.7 \\
\hline
w/o Blinding
  & 0.432 (↑0.023) & 92.3\% (↑0.1\%) & 0.832 (↑0.003)
  & 105.2 (↓3.4) & 95.1 (↓14.6) \\
w/o Permutation
  & 0.410 (↑0.001) & 97.5\% (↑5.3\%) & 0.894 (↑0.065)
  & 104.8 (↓3.8) & 83.8 (↓25.9) \\
w/o Permutation \& Blinding
  & 0.438 (↑0.029) & 97.5\% (↑5.3\%) & 0.912 (↑0.083)
  & 99.8 (↓8.8) & 70.2 (↓39.5) \\
Naive Baseline
  & 0.462 (↑0.053) & 100.0\% (↑7.8\%) & 1.000 (↑0.171)
  & 0.0 (↓108.6) & 0.0 (↓109.7) \\
\hline
\multicolumn{6}{l}{$^\dagger$ Ablation experiments use the same experimental setup as the main experiments.}
\end{tabular}%
}
\vspace{-0.5cm}
\end{table}

Evaluating cross-organizational probing, vector2Trans reduces average nDCG@10 from 0.407 to 0.037 (90.8\% reduction), achieving levels statistically indistinguishable from random retrieval (Fig.~\ref{fig:probing_attack}). This demonstrates query privacy preservation: queries transformed with one organization's key return only topically irrelevant documents from other organizations' databases, concealing query intent while maintaining operational functionality.

\subsection{Ablation Study}
\label{sec:ablation_study}

To understand how each component contributes to Trans-RAG's security-performance balance, we conducted an ablation study examining the impact of removing individual security mechanisms across our three evaluation dimensions.

\noindent\textbf{Design Rationale.} The vector2Trans transformation employs 
two complementary security mechanisms: \textbf{Permutation} creates structural 
isolation by decorrelating dimensions, preventing coordinate system recovery. 
\textbf{Cryptographic blinding} adds input-dependent noise to disrupt statistical 
patterns. Together, these balance reconstruction resistance with retrieval utility.

As shown in Table~\ref{tab:ablation_study}, our analysis reveals three key insights. First, \textbf{permutation dominates computational overhead}: removing permutation reduces processing time by 25.9ms (vs. 14.6ms when removing blinding). Second, \textbf{blinding has a larger impact on accuracy}: removing blinding improves nDCG@10 by +0.023, whereas removing permutation yields only +0.001. Third, \textbf{components exhibit complementary effects}: removing both achieves the best retrieval performance (+0.029 nDCG@10, +5.3\% overlap) and the fastest processing (-39.5ms) among ablations.

The complete Trans-RAG system achieves the optimal security-performance balance, demonstrating that both components are essential for the security properties established in Section~3: permutation for robust isolation and blinding for refined protection against statistical analysis.
\vspace{-0.3cm}
\section{Conclusion and Future Work}
Trans-RAG introduces a paradigm shift from document-level encryption to query-level transformation for secure cross-organizational RAG, implementing a vector space language framework where queries dynamically "speak" each organization's private semantic space through vector2Trans. This query-centric approach resolves the fundamental tension between security, accuracy, and efficiency: achieving near-orthogonal vector space isolation (89.90° angular separation, 99.81\% isolation) with strong attack resistance, while maintaining 96.5\% retrieval effectiveness and delivering 32,216× speedup over homomorphic encryption. By operating transparently with existing vector databases and embedding models, Trans-RAG enables practical secure collaborative knowledge sharing across organizational boundaries without infrastructure modifications, establishing a foundation for privacy-preserving vector-based retrieval in federated environments.

Future work may extend the vector space language paradigm to stronger 
adversarial settings, including malicious query manipulation, vector 
database tampering, and cross-organizational collusion, potentially 
augmented with lightweight verification or cryptographic consistency 
mechanisms. In large-scale deployments, efficient key lifecycle 
protocols for organization onboarding, revocation, and distributed 
trust establishment become increasingly important without incurring 
prohibitive system overhead. Additionally, comprehensive end-to-end 
evaluation of generation quality in full RAG pipelines and optimization 
of structured transformations for ultra-high-dimensional embeddings 
remain promising directions.

\subsubsection*{Acknowledgments.} This research is supported by the National Key R\&D Program of China (No. 2023YFC3303800).

\bibliographystyle{splncs04}
\bibliography{reference}

\end{document}